\documentclass[prb,twocolumn,superscriptaddress,showpacs]{revtex4}

\usepackage{graphicx}
\usepackage{bm}

\begin{document}

\title{Spin-density functional study of the organic polymer
    dimethylaminopyrrole: A realization of the organic periodic
    Anderson model }

\author{Yuji Suwa}
\affiliation{Advanced Research Laboratory, Hitachi, Ltd., Kokubunji,
Tokyo 185-8601, Japan}
\author{Ryotaro Arita}
\affiliation{Department of Applied Physics, University of Tokyo, Hongo,
Tokyo 113-8656, Japan}
\author{Kazuhiko Kuroki}
\affiliation{Department of Applied Physics and Chemistry,
University of Electro-Communications, Chofu, Tokyo 182-8585, Japan}
\author{Hideo Aoki}
\affiliation{Department of Physics, University of Tokyo, Hongo,
Tokyo 113-0033, Japan}

\date{\today}

\begin{abstract}
While the periodic Anderson model (PAM) has been recognized as a good
model for various heavy f-electron systems, here we design a purely
organic polymer whose low-energy physics can be captured by PAM. By
means of the spin density functional calculation, we show that polymer
of dimethylaminopyrrole is a candidate, where its ground state can
indeed be magnetic depending on the doping.  We discuss the factors
favoring ferromagnetic ground state.
\end{abstract}


\maketitle

\section{Introduction}

Realizing organic ferromagnets consisting entirely of non-magnetic
elements is a challenging target in condensed-matter physics and
chemistry\cite{Allemand,Rajca}. We usually cannot expect itinerant
ferromagnetism in organic materials with p-bands, although there are
very a few exceptions such as
polymethylthiophene\cite{Pereira,Correa,Nascimento,Paula}. Therefore,
we need a smart design/strategy to realize purely organic
ferromagnets.

Polymers of five membered rings constitute a unique and interesting
playground for such materials design, because we can attach a variety
of functional groups to the polymers. Indeed, the present authors have
proposed that we can realize ``flat-band ferromagnetism'', originally
conceived by Mielke and by Tasaki\cite{MielkeTasaki1,MielkeTasaki2},
in aminotriazole with the multi-band Hubbard
model\cite{Arita1,Arita2,Suwa1,Suwa2}. While it has been well known
that the Hubbard model can have a spin-polarized ground state, it is
also well known that the periodic Anderson model, which has been
studied extensively especially in the context of the heavy-fermion
electron systems, has various magnetic ground states that include a
metallic ferromagnet. In the present study, we explore whether a
purely organic polymer can be mapped to the periodic Anderson model.

In the course of our attempts, we have found that {\it
dimethylaminopyrrole}, which can be polymerized comparatively
easily\cite{Nishihara}, is a good candidate for that purpose in that
it can accommodate an electronic structure having both narrow and wide
bands. By means of spin-density-functional calculations, we shall
indeed show that the polymer and oligomers of dimethylaminopyrrole
have ferromagnetic (F) or antiferromagnetic (AF) instabilities if
holes are sufficiently doped (i.e., more than 1/2 hole per
five-membered ring). As for the former instability, we propose that
its origin can be explained by the scenario proposed by Batista {\it
et al.}\cite{Batista} While we find that the difference between the
energy of the F state and the AF states are rather small in this
material, we shall finally give a discussion on how the F state can be
stabilized.

\section{Dimethylaminopyrrole polymer}

\begin{figure}
\begin{center}
\includegraphics[width=6.5cm]{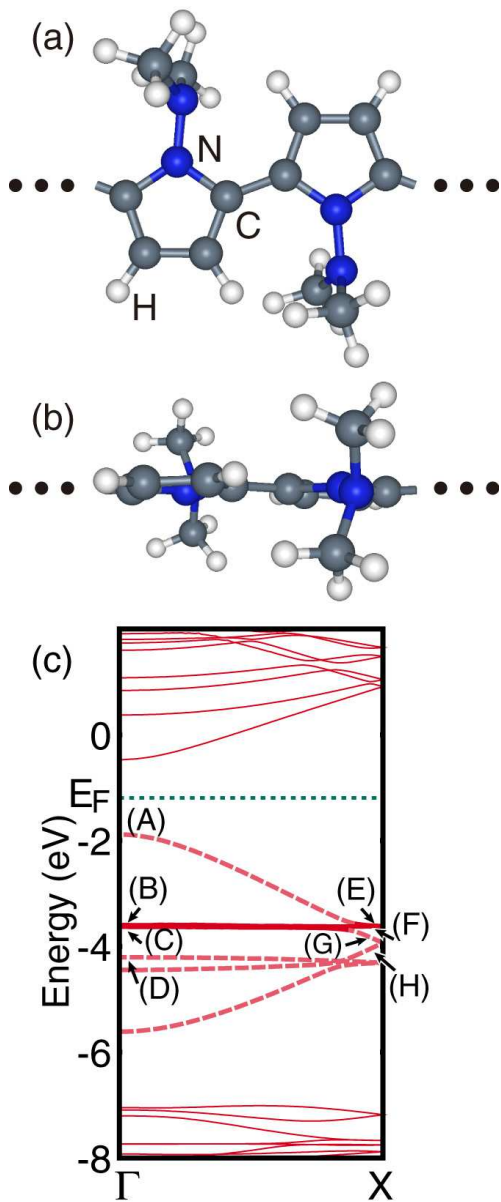}
\end{center}
\caption{(Color online) Top view (a) and side view (b) of the
 optimized atomic structure of polydimethylaminopyrrole in a unit
 cell. (c) Band structure of undoped polydimethylaminopyrrole. Thick
 solid (dashed) lines represent $\sigma$-bands ($\pi$-bands).}
 \label{XyzBand}
\end{figure}

Let us start with the undoped polymer of dimethylaminopyrrole.
Figure~\ref{XyzBand}(a,b) depicts the atomic structure of
polydimethylaminopyrrole in a unit cell.  We have studied both the
spin-unpolarized cases based on the density-functional theory with
generalized gradient approximation\cite{PBE96,White} (GGA-DFT) and the
spin-polarized ones based on the spin-density functional theory
(GGA-SDFT). We have used the plane-wave based ultrasoft
pseudopotentials\cite{Vanderbilt,Laasonen} with the energy cutoff at
20.25 Rydberg. The convergence criterion of the geometry optimization
was that all of the forces acting on each atom were within
$1\times10^{-3}$ H/a.u.  We have first optimized the structure.  While
we have started, in our materials design, from a planar configuration
of rings, the structure-optimized polydimethylaminopyrrole has
dimethylamino bases somewhat twisted out of the plane. This makes the
$\pi$-orbital of the nitrogen atom in the amino base lying in the
molecular plane, that is, the angle between the $\pi$-orbitals of the
nitrogen atom in the amino base and the neighboring nitrogen atom in
the five membered ring becomes nearly 90 degrees, leading to a very
small $\pi$-bonding between them.

The band structure of undoped polydimethylaminopyrrole is shown in
Fig.~\ref{XyzBand}(c).  We can see that there exist flat bands
(labeled as (B)-(E) and (C)-(F) in Fig.~\ref{XyzBand}(c)), but they
lie well below the Fermi energy, unlike the case of
polyaminotriazole\cite{Arita1,Arita2,Suwa1,Suwa2} where the flat band
sits around $E_F$. In the band structure, a band folding of the
quasi-one-dimensional band is due to the alternating directions of the
pyrrole molecules.

\begin{figure}
\begin{center}
\includegraphics[width=8.0cm]{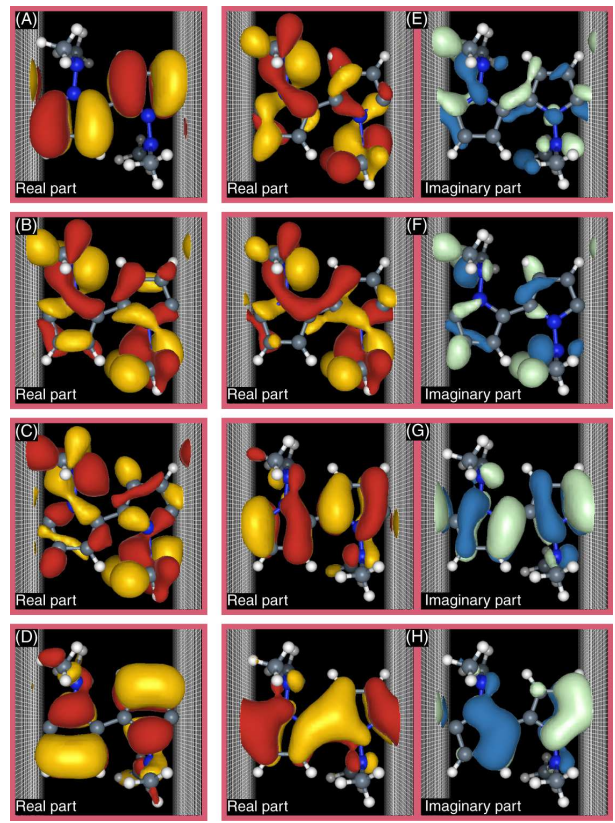}
\end{center}
\caption{(Color online) Wave functions in the highest four valence
 bands of undoped polydimethylaminopyrrole at $\Gamma$ (A-D) or at X
 (E-H) as depicted by isosurfaces with different colors representing
 the signs of the wave function.  See Fig.~\protect\ref{XyzBand}(c)
 for labels of the energy levels.  We have chosen the wave function to
 be real at $\Gamma$. }  \label{PolWfn}
\end{figure}

Figure~\ref{PolWfn}(A-H) shows the wave functions in the highest four
valence (occupied) bands at $\Gamma$ and X points. The wave functions of
the flat bands (B, C, E and F in Fig.~\ref{XyzBand}(c)) are mixtures of
the $\pi$-orbitals of the nitrogen atoms and $\sigma$-orbitals of the
carbon atoms, while the other wave functions consist almost entirely of
$\pi$-orbitals.  We can thus call the flat bands (B)-(E) and (C)-(F)
(thick solid lines in Fig.~\ref{XyzBand}(c)) as $\sigma$-bands, and the
top valence band (A)-(G) (uppermost thick dashed line in
Fig.~\ref{XyzBand}) as a $\pi$-band.

While in polyaminotriazole\cite{Suwa1,Suwa2} the $\pi$-band is flat,
for which the ``flat-band ferromagnetism'' is expected to be realized,
this is no longer the case with polydimethylaminopyrrole, since the
$\pi$-band is considerably dispersive. On the other hand the
$\sigma$-bands are flat, and one might expect that the flat-band
ferromagnetism may occur. This does not apply either, since the bands
do not satisfy the local connectivity
condition\cite{MielkeTasaki1,MielkeTasaki2} necessary for the
flat-band ferromagnetism. They are flat simply because the transfer
between five-membered rings is small, as seen in
Figs.~\ref{PolWfn}(B), (C), (E) and (F) where $\sigma$-bonds are
absent between those two carbon atoms connecting the adjacent
five-membered rings. Although small $\pi$-like orbitals can be seen
there, they are negligibly small compared to those of $\pi$-bands in
Figs.~\ref{PolWfn}(G) and (H).

\begin{table*}[tb]
\caption{Total energies (in eV) of ferromagnetic (F) and
antiferromagnetic (AF) states measured from that of nomagnetic (N)
state. The values in parenthesis indicate the differences between the
numbers of up and down spins per unit cell.  }\label{FvsAFpoly}
\begin{tabular}{c|c|c|c|c}\hline
$n_{\rm h}$  & 1.25 & 1.5 & 1.75 & 2.0 \\
\hline\hline
\raisebox{1ex}{polymer}
        & \raisebox{1ex}{F(0.17):+0.008}
        & \raisebox{1ex}{F(0.45):-0.013}
        & \shortstack[l]{F(0.71):-0.066\\ AF:-0.068}
        & \shortstack[l]{F(0.97):-0.145\\ AF:-0.152} \\
 \hline
\end{tabular}
\end{table*}

Does this imply that we have no magnetism?  If we perform GGA-SDFT
calculations of the hole-doped polydimethylaminopyrrole for various
values of hole concentration, spin polarization does appear when the
number of holes per unit cell $n_{\rm h}$ exceeds 1.0.  Spins are
ferromagnetically ordered or antifrromagnetically ordered depending on
the initial condition of the calculation. In the AF state here, up and
down spins are aligned alternately in units of a five-membered ring.

Total energies of F and AF states calculated with GGA-SDFT measured
from nonmagnetic (N) states calculated with GGA-DFT are listed in
Table~\ref{FvsAFpoly}. They are almost the same for each hole
concentration within the accuracy of the calculation. Total energies
of F and N states are also nearly the same at $n_{\rm h}=1.25$ and
1.5. F and AF states could not be obtained even as metastable states
when $n_{\rm h}$ is lower than 1.25 and 1.75, respectively.  These
results show that more than half-filled holes ($n_{\rm h}\geq 1.0$)
create magnetic instability in this system.  In the above
calculations, no Peierls instability appeared when geometry
optimization is performed for each hole concentration and spin
ordering. This fact indicates that the structure of this polymer is
quite robust.

\begin{figure*}
\begin{center}
\includegraphics[width=12cm]{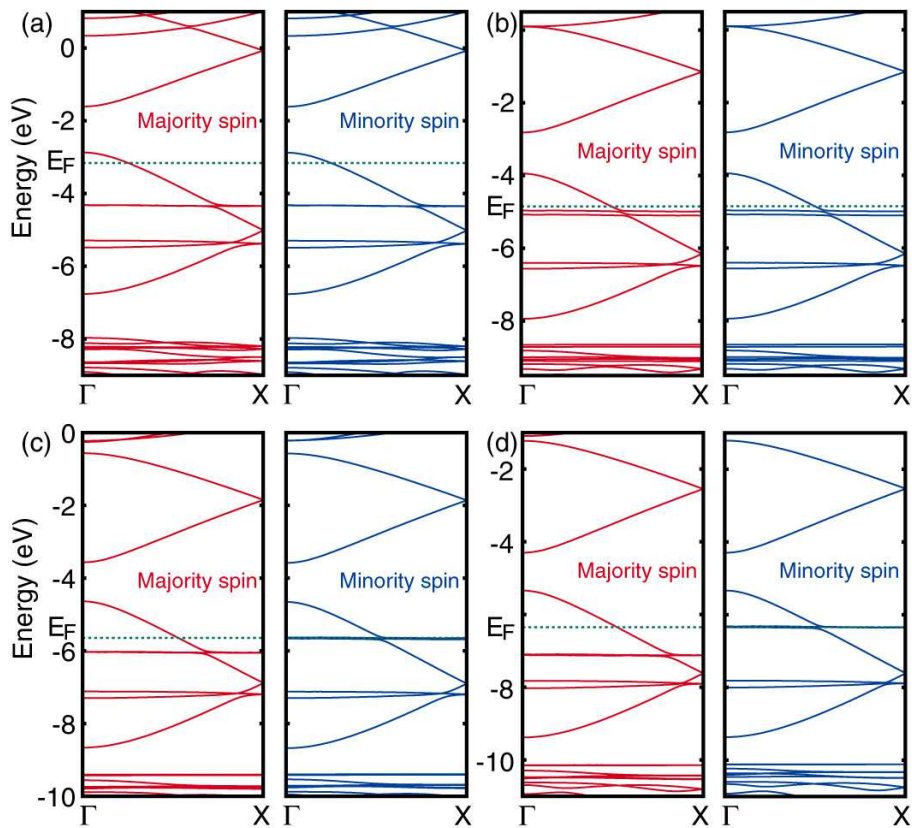}
\end{center}
\caption{(Color online) Spin-resolved band structures of the
hole-doped polydimethylaminopyrrole for the concentration of holes per
unit cell (a) $n_{\rm h}=0.5$, (b) $n_{\rm h}=1.0$, (c) $n_{\rm
h}=1.5$ (F state), (d) $n_{\rm h}=2.0$ (F state).}\label{Bands}
\end{figure*}

Fig.~\ref{Bands} shows the band structures of the hole-doped
polydimethylaminopyrrole calculated with GGA-SDFT. While there is no
spin polarization for $n_{\rm h} = 0.5$ or 1.0, spin polarization
exists in the F states at $n_{\rm h}=1.5$ and 2.0, as indicated in
Table~\ref{FvsAFpoly}.

\section{Origin of the magnetism: Periodic Anderson model}

Next let us consider in detail whether dimethylaminopyrrole can be
mapped to the periodic Anderson model, and whether it has a magnetic,
especially ferromagnetic ground state.  Indeed, here we have narrow
$\sigma$ bands embedded in wide $\pi$ bands which hybridize with each
other. Theoretically, it has been well known that the periodic
Anderson model can have a ferromagnetic ground state in a variety of
conditions\cite{Gulacsi1,Gulacsi2,Gulacsi3}. One famous example is the
situation when the model can be mapped to the Kondo lattice
model\cite{Tsunetsugu}, i.e., when there is one electron in each
localized orbital.  On the other hand, Batista {\it et
al.}\cite{Batista} recently proposed another mechanism which can work
when the particle density of the localized state is not close to
unity.

The basic idea of Batista {\it et al.}\cite{Batista} may be
recapitulated as follows.  Let us consider the periodic Anderson model
whose Hamiltonian is defined as
\begin{eqnarray}
H&=&H_0+H_U , \\
H_0&=&-t_c\sum_{r,r',\sigma}
\left(
c^\dagger_{r,\sigma}c_{r',\sigma}+{\rm h.c.}
\right) \nonumber\\
&&-t_f\sum_{r,r',\sigma}
\left(
f^\dagger_{r,\sigma}f_{r',\sigma}+{\rm h.c.}
\right)
+\varepsilon_f\sum_{r,\sigma}n_{r,\sigma}^f \nonumber\\
&&+V\sum_{r,\sigma}
\left(
c^\dagger_{r,\sigma}f_{r',\sigma}+{\rm h.c.}
\right), \\
H_U&=&-U\sum_{r}
n_{r,\uparrow}^f n_{r,\downarrow}^f ,
\end{eqnarray}
where the conduction electrons (created by $c^\dagger_{r,\sigma}$)
hybridize with localized electrons (created by $f^\dagger_{r,\sigma}$)
with $r$ representing the site and $\sigma$ the spin.  Here $t_c, t_f$
are the respective hopping energies, $\varepsilon_f$ (assumed to be
$|\varepsilon_f| < t_c$) the energy level of the localized orbital, $V$
the hybridization, and $U$ is the repulsive interaction within the
localized orbital. We can diagonalize the one-body part of the two-band
Hamiltonian $H_0$ to obtain the band representation as
\begin{equation}
H_0=\sum_{k,\sigma} 
\left(
E^+_k \beta^\dagger_{k,\sigma}\beta_{k,\sigma}
+E^-_k \alpha^\dagger_{k,\sigma}\alpha_{k,\sigma}
\right), \label{EPlusMinus}
\end{equation}
where $E^+_k > E^-_k$ (see Fig.\ref{UnfoldedBands}).

Let us introduce two kinds of Wannier orbitals, 
one dominantly localized,
\begin{eqnarray}
\phi_{r,\sigma}&=& \frac{1}{N}\sum_k^{K^{>}} 
e^{{\rm i} kr} \beta_{k,\sigma}
+\frac{1}{N}\sum_k^{K^{<}} 
e^{{\rm i} kr} \alpha_{k,\sigma},
\end{eqnarray}
and another dominantly conductive,
\begin{eqnarray}
\psi_{r,\sigma}&=& \frac{1}{N}\sum_k^{K^{<}} 
e^{{\rm i} kr} \alpha_{k,\sigma}
+\frac{1}{N}\sum_k^{K^{<}} 
e^{{\rm i} kr} \beta_{k,\sigma},
\label{PhiPsi}
\end{eqnarray}
where we have decomposed the $k$-space into $K^{>}$ ($K^{<}$) for which
the localized (conduction) electrons are the dominant in the
$\beta_{k,\sigma}$ band (Fig.\ref{UnfoldedBands}).

With the new Wannier basis the one-body part of the Hamiltonian now
reads
\begin{equation}
H_0=\sum_{r,r'}\tau^{\psi}_{r,r'}
\psi^\dagger_{r,\sigma}\psi_{r',\sigma}
+\tau^{\phi}_{r,r'}
\phi^\dagger_{r,\sigma}\phi_{r',\sigma},
\end{equation}
where $\tau$'s are new transfer energies with no off-diagonal terms
like $\psi^\dagger_{r,\sigma}\phi_{r',\sigma}$.  Batista {\it et
al.}\cite{Batista} have shown that, while the interaction term in the
Hamiltonian becomes complicated in terms of $\psi$ and $\phi$, the
dominant term is $U_{\rm eff}n_{r\sigma}^\phi n_{r}^\phi$, while the
other terms can be neglected when the hybridization $V$ is not too
strong.  When the system is in the {\it mixed valence} regime, namely,
when the Fermi energy $E_F$ is above $\varepsilon_f$ in the present
hole-doped case, holes first doubly-occupy the uncorrelated $\psi_k$
states which are above $\varepsilon_f$, and then go into $\phi_k$
states. There $\phi_k$'s tend to have parallel spins, because the cost
of the kinetic energy due to the spin polarization ($\sim t_f$) is
small, while the ferromagnetic state does not feel $U_{\rm eff}$
because of the Pauli's exclusion principle.

When all the $\phi_k$'s are singly occupied (by holes in the present
case), we can construct localized $\phi_r$'s by using occupied states
only. Because there is no interaction between the spins in this case,
the ferromagnetic state and the paramagnetic state have the same
energy. When $\phi_k$'s are only partly occupied, by contrast, we
cannot construct completely localized $\phi_r$'s only from the
occupied states. Carriers then start to feel $U_{\rm eff}$ through the
overlap of delocalized $\phi_r$'s, which makes the ferromagnetic state
energetically favorable. Therefore, the system should choose a
ferromagnetic ground state when $\phi_k$ states are less than
half-filled for $U_{\rm eff}$ sufficiently greater than $t_f$.  This
has been demonstrated numerically for the one-dimensional periodic
Anderson model.\cite{Batista}

In this mechanism of ferromagnetism, a hybridization between narrow and
dispersive bands plays an important role. In polydimethylaminopyrrole,
it is clear from the band structures in Fig.~\ref{XyzBand}(c) and
Fig.\ref{Bands} (which are summarized in Fig.\ref{UnfoldedBands}) the
$\sigma$-band and $\pi$-band are hybridized each other. Although the two
bands may seem intersecting, they are actually anticrossing with each
other.

\section{Tight-binding analysis}

\begin{figure}
\begin{center}
\includegraphics[width=8cm]{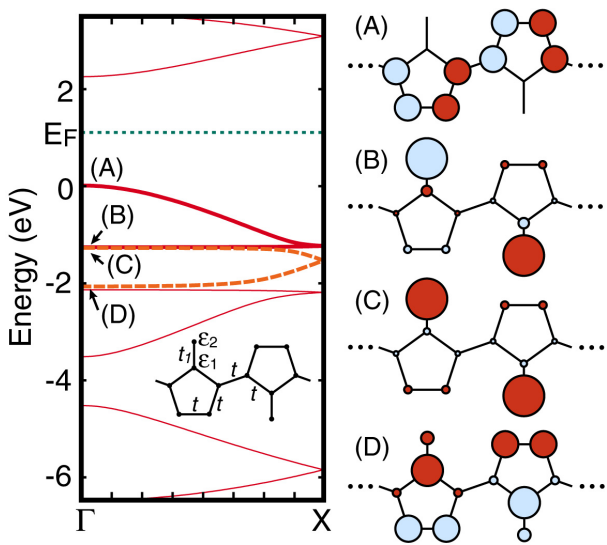}
\end{center}
\caption{(Color online) Left: Band structure of
polydimethylaminopyrrole in a tight-binding model. Inset shows
tight-binding parameters. They are fitted to reproduce
Fig.~\protect\ref{XyzBand}(c). Thick solid (dashed) lines correspond
to $E^+_k$ ($E^-_k$) in Eq.~(\protect\ref{EPlusMinus}).  Right: Wave
functions for the highest four occupied states (A-D labeled in the
left panel) at $\Gamma$, where the size of each circle represents the
amplitude while the different colors its sign.} \label{TbBands}
\end{figure}

Let us then construct the one-body part of the periodic Anderson
Hamiltonian from first principles. The model consists of five sites
representing the $\pi$-orbitals in the five-membered ring and an
additional ``amino N'' site.  The amino N site represents the
molecular orbital that consists of the $\pi$-orbital of the N atom in
the amino base, the $\sigma$ (sp$^3$) orbitals of two methyl bases and
$\sigma$-orbitals of the five-membered ring. The transfer and on-site
energies shown in the inset of Fig.~\ref{TbBands} are fitted to
reproduce Fig.~\protect\ref{XyzBand}(c). Obtained values are: $t=2.5$
eV, $t_1=2.0$ eV, $\varepsilon_1=-1.4$ eV, and $\varepsilon_2=-1.3$
eV. Figure~\ref{TbBands} shows the band structure and wave functions
in the tight-binding model. In the tight-binding model the primitive
cell contains only one five-membered ring, but we have folded the
bands at X ($k=\pi/2a$) to facilitate comparison with Fig.1(c). We can
see that the tight-binding fit successfully reproduces all the
relevant (i.e., the thick lines in the DFT result in Fig.1(c))
$\pi$-bands and $\sigma$-band. So the six sites considered in the
tight-binding model have turned out to be sufficient for describing
the relevant bands (consisting of the mainly $\pi$ bands of the ring,
plus another one whose character is not the $\pi$ orbits of the
ring). In other words, the bonding between the ring and amino N can be
expressed only by the transfer between the $\pi$-orbital of the N atom
at the apex of the ring (apex N) and that at the amino N. Because the
molecular orbital that forms $\sigma$-band is included in the amino N
site, the $\sigma$-band of this model has a large amplitude at the
amino N.  In this sense, ``$\sigma$-band'' (which was meant to mean
those other than the $\pi$ orbits of the ring) should rather be called
``dimethylamino band'', or, put more generally, the functional group
band.

The simple model shows that the $\sigma$-band does not satisfy local
connectivity condition. The flatness of the band simply comes from
localized nature of the wave function. While the tight-binding model
here is the same as that for polyaminotriazole\cite{Suwa1} except for
the values of parameters, a large difference is that the transfer
between the amino N and apex N is very small here compared to that in
polyaminotriazole, which is precisely because the $\pi$-orbitals of the
N atoms are twisted almost 90 degrees with each other here. The twist
also produces hybridization with $\sigma$-orbitals of the five-membered
ring. The small transfer between the amino N and apex N breaks the
subtle balance of parameters necessary to satisfy local connectivity
condition for the flat-band ferromagnetism, but simultaneously creates
narrow and dispersive bands, mutually hybridized, which is a
prerequisite for the ferromagnetism based on the periodic Anderson
model.

\begin{figure}
\begin{center}
\includegraphics[width=7.5cm]{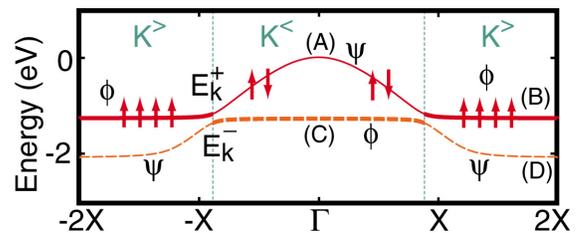}
\end{center}
\caption{(Color online) Band structure of polydimethylaminopyrrole in
the tight-binding model, unfolded onto an extended Brillouin
zone. Solid (dashed) lines correspond to $E^+_k$ ($E^-_k$) in
Eq.~(\protect\ref{EPlusMinus}), while thick (thin) lines $\phi$
($\psi$) states in Eq.~(\protect\ref{PhiPsi}) ($\sim \sigma$ ($\pi$)
bands in the present system).  Labels (A-D) refer to the band
structure in Fig.~\protect\ref{TbBands}.  Arrows schematically
indicate the spins of accommodated holes.  }
\label{UnfoldedBands}
\end{figure}

In order to make the identification of the present band structure with
the periodic Anderson model clearer in terms of $\phi$ and $\psi$
states in Eq.~(\ref{PhiPsi}), we show in Fig.~\ref{UnfoldedBands} the
band structure (Fig.~\ref{TbBands}) unfolded into an extended
Brillouin zone. The upper band (solid line) consisting of a part of
the top valence $\pi$-band and a part of the flat $\sigma$-band
correspond to $E^+_k$ in Eq.~(\ref{EPlusMinus}), while the lower band
(dashed line) consisting of the other parts of $\sigma$ and
$\pi$-bands correspond to $E^-_k$.  In the figure, thick lines and
thin lines correspond to $\phi$ and $\psi$-states in
Eq.~(\ref{PhiPsi}), respectively.  Simply put, $\phi$ represents the
$\sigma$-band and $\psi$ the $\pi$-band in the present organic system.

By using obtained tight-binding parameters of the two bands, we
performed diagonalization study of periodic Anderson model where each
site represents a five-membered ring and six sites form a ring.
Although this model is essentially the same as
Ref.~[\onlinecite{Batista}] except for the values of parameters,
obtained ground state was not ferromagnetic state. This result is
compatible with the result of GGA-SDFT where total energy of F state
is not definitely lower than that of AF state. This is probably due to
the small hybridization $V$ (within the range $V/t_c=0.01$ to 0.1)
compared to that in Ref.~[\onlinecite{Batista}] (typically $V/t_c=0.3$
or 0.5).

In terms of the tight-binding model, we can explain an interesting
feature in the GGA-SDFT band structure (Fig.~\ref{Bands}) where the
energy of the $\sigma$-band shifts relative to the other bands with
the doping.  For instance, as we increase the number of holes from
$n_{\rm h}=0$ (Fig.~\ref{Bands}(a)) to 1.0 (Fig.~\ref{Bands}(b)), the
$\sigma$-band goes up. When $n_{\rm h}$ is increased further
(Figs.~\ref{Bands}(c) and \ref{Bands}(d)) , the $\sigma$-band of
majority spins starts to go down, while that of minority spins does
not.  As a result, the lower limit of doping concentration necessary
for the spin polarization becomes nearly half-filled ($n_{\rm h}=1.0$,
i.e., 1/2 hole per five membered ring). Such a shift of the
$\sigma$-band in the GGA-SDFT calculation can be reproduced by using
the tight-binding model, if a Coulomb interaction of 1.4 eV between
the amino N site and five-membered ring is assumed and a mean-field
approximation is applied. Note that $\pi$ orbital of the amino N is
hybridized with the $\sigma$ orbitals of the five-membered ring, so
that the interaction effectively contains an intra-site interaction
and can be large.

\section{Oligomers}

We move on to the oligomers of dimethylaminopyrrole. This is
practically important because synthesis of the oligomers of
dimethylaminopyrrole is possible\cite{Nishihara} while the polymer may
be difficult to fabricate. Like the polymer, oligomers of
dimethylaminopyrrole have spin polarization if holes are doped
enough. Total energies of F state and AF state are almost the same for
each length and number of holes, as will be shown later.

\begin{figure}
\begin{center}
\includegraphics[width=7cm]{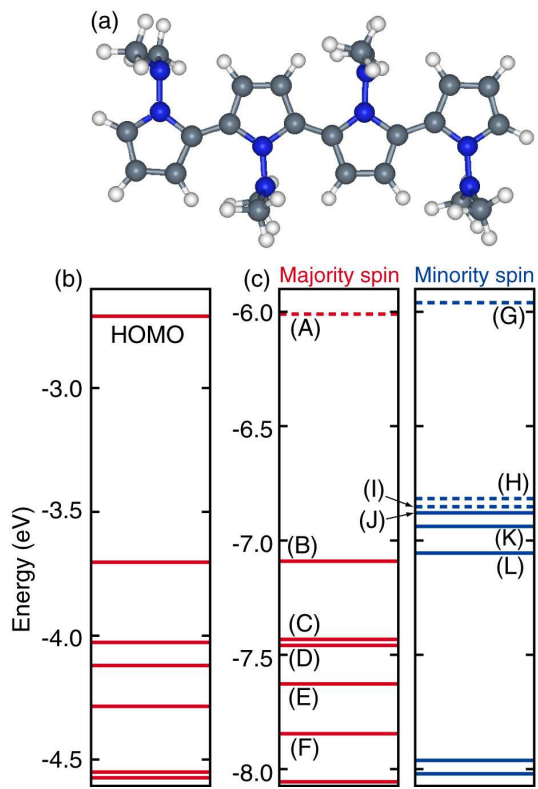}
\end{center}
\caption{(Color online) (a) Atomic structure and energy levels of (b)
undoped and (c) four-hole-doped (F state) tetramer of
dimethylaminopyrrole.  Solid (dashed) lines represent occupied
(unoccupied) states.  Wave functions of the states labeled as (A-L)
are shown in Fig.~\protect\ref{OliWfn} below. }
\label{Levels}
\end{figure}

As a typical oligomer, we show the result for a tetramer.
Figure~\ref{Levels} shows the atomic structure, the energy levels of
undoped nonmagnetic tetramer and four-hole-doped ferromagnetic
tetramer.  For the ferromagnetic tetramer, we can see in
Fig.~\ref{Levels}(c) that four consecutive minority-spin states (H-K)
take relatively higher energy levels than corresponding majority-spin
states (C-F), while other levels are almost the same between the
majority and minority spins.  As a consequence the higher two states
out of the four become unoccupied in the minority-spin channel.

\begin{figure}
\begin{center}
\includegraphics[width=7.5cm]{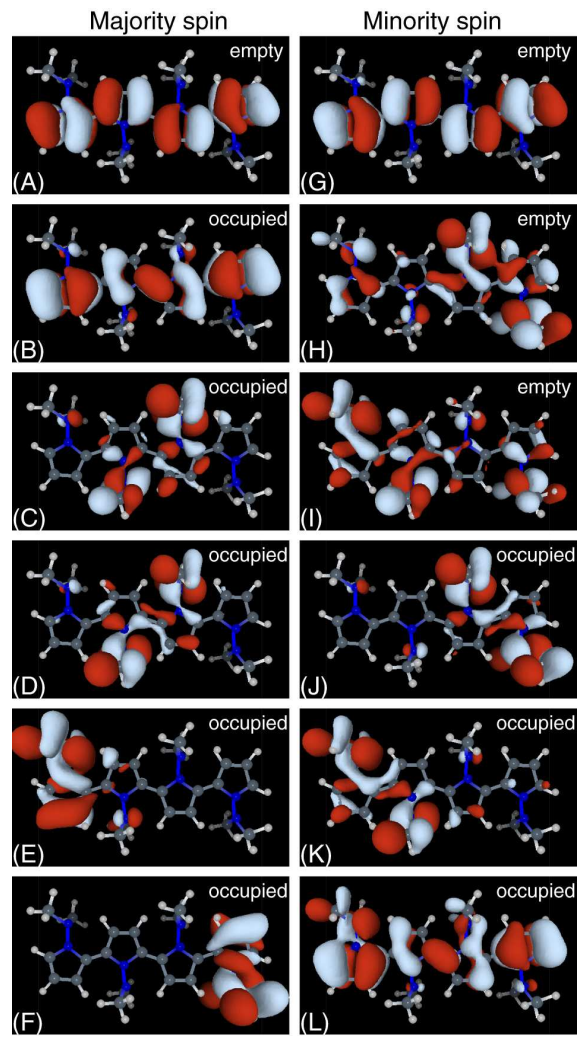}
\end{center}
\caption{(Color online) Wave functions of a four-hole-doped tetramer
of dimethylaminopyrrole for the energy levels labeled in
Fig.~\protect\ref{Levels}(c) as depicted by isosurfaces with different
colors representing the signs of the wave function.}  \label{OliWfn}
\end{figure}

The four states are $\sigma$-states, counterparts of the $\sigma$-band
in the polymer, which can be confirmed if we look at the wave functions
of the four states in Figs.~\ref{OliWfn}(C-F) for majority spin and in
Figs.~\ref{OliWfn}(H-K) for minority spin. All these wave functions are
seen to consist of nitrogen $\pi$-orbitals and carbon $\sigma$-orbitals,
while the other wave functions in Figs.~\ref{OliWfn}(A,B,G,L) consist
only of $\pi$-orbitals.

We can relate the arrangement of energy levels of the oligomer with the
band structure of the polymer. Generally, for an oligomer consisting of
$n$ monomers ($n$-mer), the number of electronic states constructed from
one kind of molecular orbital of the monomer is $n$. Obviously, a state
in which all monomers have the same phase is one of the $n$ states, and
its wave function is similar to that of the polymer at $\Gamma$-point
except for the truncation of the chain. We call it $\Gamma$-state.
Another linearly independent state can be constructed by introducing a
node into the wave function of the $\Gamma$-state making the signs of
the molecular orbitals at the opposite ends reversed. This state is
similar to that of the polymer at $k=\pi/na$ ($a$: the length of the
monomer) in that the phase rotates by $\pi$ per $n$ monomers. In the
same way, $m$-th state of the oligomer corresponding to the polymer's
state at $k=(m-1)\pi/na$ can be constructed by introducing $m-1$ nodes
(a phase rotation of $(m-1)\pi$) in the $\Gamma$-state. Because of the
similarity of the wave functions, the energy level of the $m$-th state
in $n$-mer can be inferred according to the band energy of the polymer
at $k=(m-1)\pi/na$.

\begin{figure}
\begin{center}
\includegraphics[width=7cm]{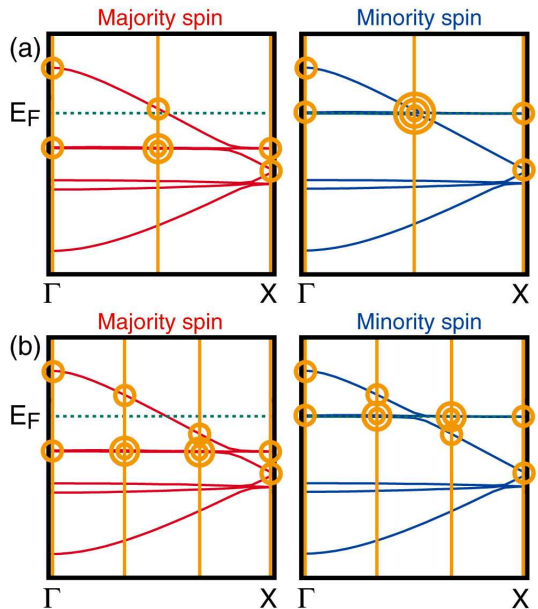}
\end{center}
\caption{(Color online) Correspondence between the band structure of
the polymer (curves) and the energy levels of oligomers (circles) for
dimethylaminopyrrole for the ferromagnetic (a) four-hole-doped tetramer
and (b) six-hole-doped hexamer. Single, double, and triple circles
indicate one, two and three energy levels in the oligomer,
respectively.}
\label{Dividing}
\end{figure}

An easy way to do such an estimation is to take the intersecting
points of the band with vertical lines dividing the zone in k-space
($k=0$ to $\pi/a$) into $n$. For example, the energy levels of
ferromagnetic four-hole-doped tetramer and six-hole-doped hexamer
around the LUMO-HOMO can be estimated as in Figs.~\ref{Dividing}(a)
and \ref{Dividing}(b), respectively.  Note again that the whole band
is folded at X ($k=\pi/2a$) because of two monomers in the unit
cell. The band structure of Fig.~\ref{Bands}(d) is used here because
the hole density of these examples is $n_{\rm h}=2.0$. We take $n$ to
be even in the following for simplicity. As shown in the following,
the order of energy levels estimated in this way agrees well with the
result for the oligomers.  One reason why such an estimation is valid
is that the finite-size effect is not very large in the oligomers of
dimethylaminopyrrole, probably due to the simple chain structure and
not-too-large transfer between monomers.

With such a kind of level scheme, we can estimate how many holes
should be doped to realize spin polarization in a given oligomer.  For
that purpose, whether the vertical line in Fig.~\ref{Dividing} exist
at $k=$X/2 is important, since the $\pi$-band and the flat
$\sigma$-band of minority spin cross at this point.  When $n/2$ is odd
as in Fig.~\ref{Dividing}(b), no vertical line exists at X/2 and thus
$n/2+1$ holes first enter the levels coming from the $\pi$-band whose
energies are the same for both spins.  Since two more holes are
necessary to occupy the levels associated with the flat $\sigma$-band,
$n/2+3$ holes are necessary in total for the state to be
spin-polarized.  The total spin in F state (i.e., the difference of
the number of up and down spins, $N_{\uparrow}-N_{\downarrow}$) can be
estimated to be $N-n/2-1$, where $N$ is the number of doped holes in
an $n$-mer.  When $n/2$ is even, on the other hand, one of the
vertical lines passes through X/2 as in Fig.~\ref{Dividing}(a). At
$k=X/2$, the four energy levels, two from the $\pi$-bands for both
spins and two from the $\sigma$-bands for the minority spin, almost
coincide when holes are doped. Depending on which level is the
highest, $n/2+2$ or $n/2+4$ holes are necessary to have
spin-polarizations in this case, with $N_{\uparrow}-N_{\downarrow}$
estimated to be $N-n/2$ or $N-n/2-2$, respectively.

\begin{table*}[tb]
\caption{Total energies (in eV) of F and AF states measured from N
state are shown for various oligomers (rows) with various numbers of
holes (columns).  Boxes marked with N denote that the total energies
of F and AF states are higher than that of N state or do not give
converged results.  The values in parenthesis indicate
$N_{\uparrow}-N_{\downarrow}$ for F state. }\label{Nmers}
\begin{tabular}{c|c|c|c|c|c|c}\hline
   & 2 holes & 4 holes & 6 holes & 8 holes & 10 holes & 12 holes \\
\hline\hline
2-mer&N& & & & & \\
\raisebox{1ex}{4-mer}&\raisebox{1ex}{N}&
       \shortstack[l]{F(2.00):-0.18 \\ AF:-0.20}
       & & & & \\
\raisebox{1ex}{6-mer}&\raisebox{1ex}{N}&\raisebox{1ex}{N}&
       \shortstack[l]{F(2.24):-0.20 \\ AF:-0.19}
       & & & \\
\raisebox{1ex}{8-mer}&\raisebox{1ex}{N}&\raisebox{1ex}{N}&
       \shortstack[l]{F(1.84):-0.07 \\ AF:-0.06} &
       \shortstack[l]{F(3.89):-0.39 \\ AF:-0.36}
        & & \\
\raisebox{1ex}{10-mer}&\raisebox{1ex}{N}&\raisebox{1ex}{N}&
       \raisebox{1ex}{N}&
       \shortstack[l]{F(2.63):-0.08 \\ AF:-0.08} &
       \shortstack[l]{F(4.11):-0.51 \\ AF:-0.50} & \\
\raisebox{1ex}{12-mer}&\raisebox{1ex}{N}&\raisebox{1ex}{N}&
       \raisebox{1ex}{N}&
       \shortstack[l]{F(1.84):-0.01 \\ AF:0.00} &
       \shortstack[l]{F(3.70):-0.21 \\ AF:-0.22} &
       \shortstack[l]{F(5.55):-0.57 \\ AF:-0.56} \\
\hline
\end{tabular}
\end{table*}

Table~\ref{Nmers} shows the result of GGA-SDFT calculations for a
series of oligomers of dimethylaminopyrrole with various numbers of
doped holes. In the calculation of the dimer and tetramer, geometry
optimizations are performed for each hole concentration and each spin
state, N, F and AF. For the hexamer and larger oligomers, geometry
optimization is performed only for undoped N state. Such a treatment
is acceptable because there is no significant difference in the
optimized structures depending on the spin state and hole
concentration in the polymers and small oligomers. The numbers of
holes necessary for spin polarization and
$N_{\uparrow}-N_{\downarrow}$ of F state almost fully agree with the
above estimate in terms of the polymer's band structure. For the
$n$-mers with odd $n/2$, i.e., hexamer and decamer, the values of the
spin polarization agree well with the above estimate, which should
reflect the fact that no energy levels associated with the $\pi$-band
exist near the Fermi energy when spin polarization emerges, as shown
in Fig.~\ref{Dividing}(b). For the oligomers with even $n/2$, i.e.,
octamer and dodecamer, $N_{\uparrow}-N_{\downarrow}$ is expected to
take either of $N-n/2$ and $N-n/2-2$ depending on the order of almost
degenerated energy levels of $\pi$-band and $\sigma$-band. The result
shows that $N_{\uparrow}-N_{\downarrow}$ is always equal to $N-n/2$
for all cases. This indicates that the energy level of the $\pi$-band
at X/2 is slightly higher than that of $\sigma$-band.

\section{Conclusion}

We have used a first-principles calculation to examine magnetism in a
polymer and oligomers of dimethylaminopyrrole. We have found that
polydimethylaminopyrrole should have magnetic instability if density
of doped holes are more than one hole per unit cell ($n_{\rm h}>1.0$).
For $n$-mers of dimethylaminopyrrole, we conclude that $n/2+2$ or
$n/2+3$ holes are necessary to have spin-polarization depending on
whether $n/2$ is even or odd, respectively.  While
polydimethylaminopyrrole belongs to the same category (chains of
five-membered rings) as the previously proposed
polyaminotriazole\cite{Arita1,Suwa1} which has theoretically shown to
realize the flat-band
ferromagnetism\cite{MielkeTasaki1,MielkeTasaki2}, we suggest that the
present magnetic instability can be related to the mechanism proposed
by Batista {\it et al.}\cite{Batista} based on the periodic Anderson
model.

Realization of ferromagnetism is not yet sufficient, because the total
energies of the ferromagnetic state and antiferromagnetic state is
almost degenerate in GGA-SDFT calculations. Diagonalizations of the
model Hamiltonians constructed from the band structure do not show
ferromagnetic ground states, either. Since this comes from the
smallness of the hybridization $V$ between the flat band and the
dispersive band, we can conversely predict that ferromagnetism will be
stabilized if we can increase $V$.  In the present material $V$ is
determined mainly by $\pi$-bonding between the nitrogen atom in the
amino base and the nitrogen atom in the five membered ring.  In
dimethylaminopyrrole, $V$ is very small because the angle between the
two $\pi$-orbitals is nearly 90 degrees. Therefore, $V$ can be
increased if this angle is decreased by replacing dimethylamino base
with another base.  Note that reducing the angle to 0 would be
excessive because that is the same as the case of aminotriazole whose
electronic state applies to the mechanism of flat-band ferromagnetism
rather than periodic Anderson model.

So the present work gives an interesting case of material design where
control of the parameters of electronic states has lead to a different
mechanism for the many-body effects, namely {\it organic periodic
Anderson model} and associated ferromagnetism. An important future
problem is whether the SDFT calculations can capture the
ferromagnetism {\it \`{a} la} Batista.  Another future problem is how
the present concept of the organic periodic Anderson model can be
extended to wider materials beyond those considered here. Studying
inter-oligomer couplings will also be important for examining
ferromagnetic materials. As for the doping the material, we can
consider various chemical dopants as discussed for in
Ref.~[\onlinecite{Arita2}].

\begin{acknowledgments}
We would like to thank Hiroshi Nishihara, Yoshinori Yamanoi and Norikazu
Ohshima for extensive discussions on the chemistry of polymers and
oligomers, and also for the collaboration in fabricating the materials.
This work was partly supported by a Grant-in-Aid for Scientific Research
from Japan Society for the Promotion of Science.  The first-principles
calculations were performed with TAPP (the Tokyo Ab-initio Program
Package).
\end{acknowledgments}

\end{document}